\documentclass[%
 aip,
 apl,
 amsmath,amssymb,
 floatfix,
preprint,%
]{revtex4-1}

\usepackage[dvips]{graphicx}% Include figure files
\usepackage{dcolumn}% Align table columns on decimal point
\usepackage{bm}% bold math
\usepackage{amsmath}
\newcommand{\angstrom}{ \text{\normalfont\AA}  }
\usepackage{xcolor}
\usepackage{float}
\usepackage{multirow}
\usepackage[utf8]{inputenc}
\usepackage[T1]{fontenc}
\usepackage{mathptmx}
\graphicspath{{./Figures/}}

\begin{document}

\title{
Resolving Multiphoton Processes with High-Order Anisotropy Ultrafast X-ray Scattering
}

\author{Adi Natan}
 \email{natan@stanford.edu}
\affiliation{Stanford PULSE Institute, SLAC National Accelerator Laboratory\\
2575 Sand Hill Road, Menlo Park, CA 94025}

\author{Aviad Schori}
\affiliation{Stanford PULSE Institute, SLAC National Accelerator Laboratory\\
	2575 Sand Hill Road, Menlo Park, CA 94025}

\author{Grace Owolabi}
\affiliation{Department Electrical Engineering and Computer Science, Howard University, Washington DC 20059, United States of America}

\author{James P. Cryan}
\affiliation{Stanford PULSE Institute, SLAC National Accelerator Laboratory\\
	2575 Sand Hill Road, Menlo Park, CA 94025}
\affiliation{Linac Coherent Lightsource, SLAC National Accelerator Laboratory, 2575 Sand Hill Road, Menlo Park, CA 94025, USA}

\author{James M. Glownia}
\affiliation{Linac Coherent Lightsource, SLAC National Accelerator Laboratory, 2575 Sand Hill Road, Menlo Park, CA 94025, USA}
 
\author{Philip H. Bucksbaum}
\affiliation{Stanford PULSE Institute, SLAC National Accelerator Laboratory\\
2575 Sand Hill Road, Menlo Park, CA 94025}
\affiliation{Department of Physics, Stanford University, Stanford, CA 94305}
\affiliation{Department of Applied Physics, Stanford University, Stanford, CA 94305}

\date{\today}

\begin{abstract}
We present first results on ultrafast X-ray scattering of strongly driven molecular Iodine and analysis of high-order anisotropic components of the scattering signal, up to four-photon absorption. We discuss the technical details of retrieving high fidelity high-order anisotropy components, and outline a method to analyze the scattering signal using Legendre decomposition. We use simulated anisotropic scattering signals and Fourier analysis to map how anisotropic dissociation motions can be extracted from the various Legendre orders. We observe multitude dissociation and vibration motions simultaneously arising from various multiphoton transitions. We use the anisotropy information of the scattering signal to disentangle the different processes and assign their dissociation velocities  on the Angstrom and femtosecond scales de-novo.
\end{abstract}

\maketitle

\begin{figure}[hbt!]
	\includegraphics[width=0.5\textwidth]{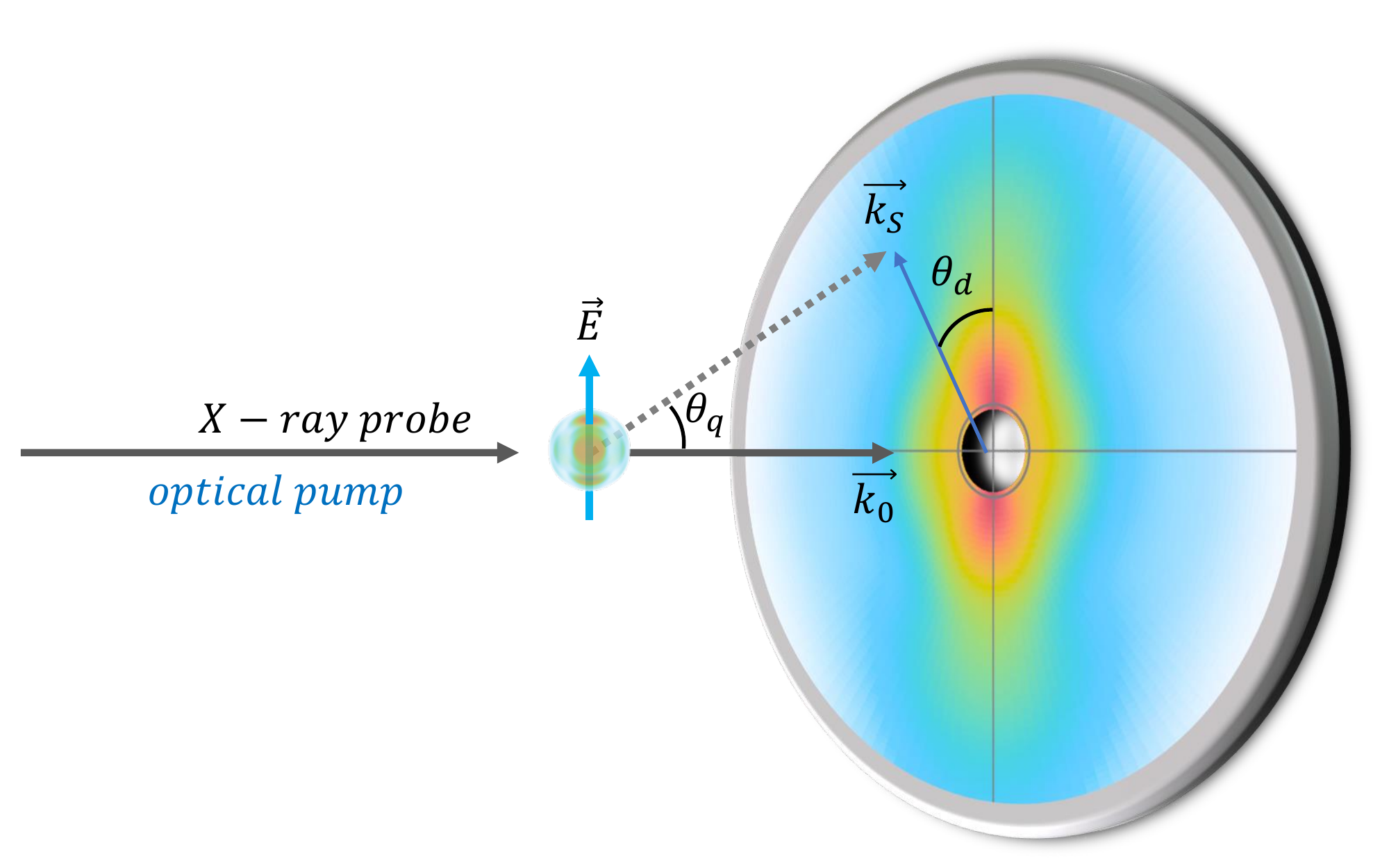}
	\caption{Schematic description of a pump-probe experimental setup and relevant coordinates presented in the text. A linearly polarized ultrashort optical laser pulse pumps the molecular sample. The sample is then probed at some time delay by an ultrashort X-ray pulse via scattering on a 2D detector array with a finite range of scattering angles. The anisotropy induced in the sample following photoexcitation is manifested by the anisotropy of the scattering pattern.}
	\label{fig:Scheme}
\end{figure}

%%%MAIN TEXT%%%%
Capturing motions in space and time at the atomic scale is fundamental to the understanding of chemical reactions and structural dynamics of molecules of different complexities.   Some of the emerging tools that allow such studies are ultrafast scattering modalities, primarily by X-rays and relativistic electrons, that were made feasible in recent years. In these studies, the motions from excited molecules are captured in a pump-probe scheme, where the excitation pump pulse is often an ultra-short linearly polarized optical laser pulse with a duration shorter than the typical timescales of motion of interest, as illustrated in Fig \ref{fig:Scheme}.  The scattering signal is then usually integrated over angle for improved fidelity and subtracted from the scattering signal of the unexcited system, to allow tracing changes of signal positions and infer structural dynamics \cite{ihee2010ultrafast}.  This approach was used successfully to demonstrate coherent motions and dynamics in molecules in the gas phase \cite{minitti2015imaging,stankus2019ultrafast}, as well as structural changes in molecules in solution after the electronic excitation \cite{biasin2016femtosecond,van2016atomistic,Chollet}. In these experiments, the optical pulse parameters were carefully chosen to ensure that only single-photon absorption is taking place and that the molecular system is photoexcited to a specific electronic state. This is often done by varying the pump intensity in the pump-probe setup and finding conditions where linearity of the excitation, as manifested by the scattering signal is achieved. The motion is inferred by measuring the scattering difference as a function of delay and applying modeling and simulations.  Using angle integrated scattering signal is well justified as it captures all types of motions that take place in the photoexcited system. However, angle-dependent signals can be dramatically attenuated if only the isotropic component is being analyzed.  In many cases, there is an inherent anisotropy in the scattering signal when a sample is excited by linearly polarized light due to an optically induced dipole moment transition. This interaction creates geometric alignment in the ensemble, and can be used to filter and enhance the specific processes under study, such as in the case of a single-photon absorption process \cite{glownia2016self,biasin2018anisotropy,haldrup2019ultrafast}, as well as perturbative two-photon excitation \cite{baskin2006oriented,yang2018imaging}.

Higher orders of anisotropy play a significant role in understanding and probing cases where the molecular system is in the presence of multi-photon absorption and strong laser fields, such as dissociation due to bond softening \cite{bucksbaum1990softening}, above-threshold dissociation \cite{mckenna2008enhancing}, quantum coherent control \cite{natan2012quantum},  and light-induced conical intersections \cite{natan2016observation}. In addition, the interaction of ultrashort pulses with molecules with anisotropic polarizability will generate  non-adiabatic (or impulsive) alignment \cite{rosca2001experimental}. The broad bandwidth of an ultrashort pulse creates rotational wavepackets that evolve and rephase at periodic time delays, forming molecular alignment, manifested by high order anisotropy in the sample under field-free conditions. Molecular alignment is often used to probe diverse phenomena in the molecular frame, such as, polyatomic vibrational dynamics and fragmentation \cite{bisgaard2009time}, ultrafast molecular frame electronic coherences  \cite{makhija2020ultrafast}, laser-induced rotational dynamics and control \cite{karamatskos2019molecular,larsen1999controlling},  Auger decay of double core-hole states\cite{cryan2010auger},  high-harmonic generation from inner valence orbitals \cite{mcfarland2008high} and diffractive imaging in the molecular frame \cite{yang2015imaging,yang2016diffractive,kierspel2020x,xiong2020high,kierspel2015strongly}.

\begin{figure}[hbt!]
	\includegraphics[width=0.5\textwidth]{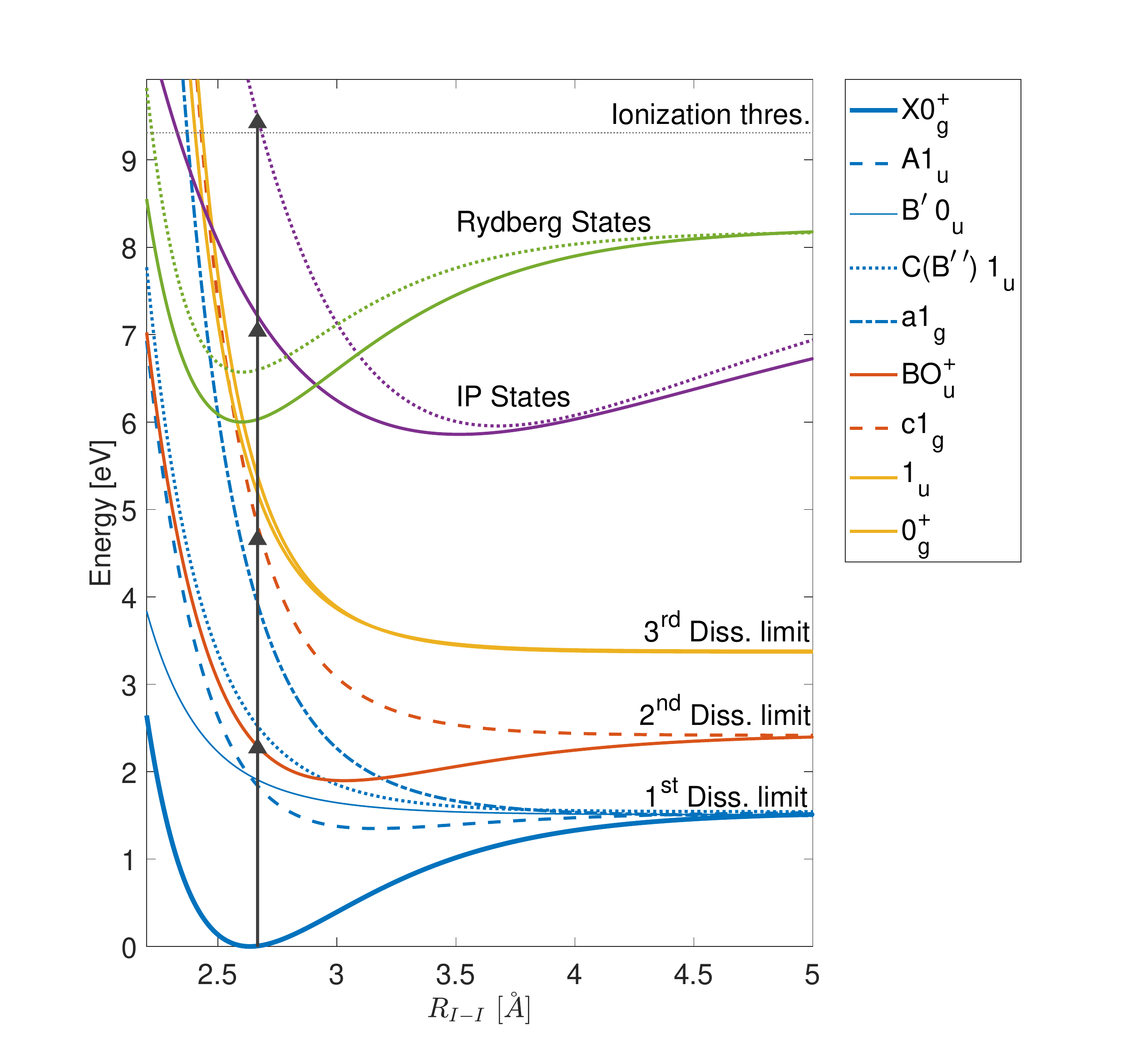}
	\caption{Potential energy curves of several valence, Rydberg and ion-pair (IP) states  of the Iodine molecule \cite{lukashov2018iodine}. Arrows represent the energy of the laser excitation (520 nm) for the case of single and multi-photon absorption.}
	\label{fig:I2curves}
\end{figure}

For the case of an ensemble of diatomic molecular Iodine that will be discussed in this work, multi-photon absorption will excite wavepackets from its ground state to a multitude of states and pathways \cite{brown1999population,pastirk2001femtosecond,fang2008strong}, including three dissociation limits, as well as higher bound Rydberg and ion-pair (IP) states, with crossing occurring between them \cite{bogomolov2014predissociation,lukashov2018iodine}.
Such excitation will create time-dependent angular distributions that will carry information regarding the number of photons absorbed, the symmetry of the states involved,  and the type of motion that is taking place.  Fig \ref{fig:I2curves} describes some of the relevant potential curves and states mentioned.

In a single-photon absorption ($\lambda=520 nm$) process, the dynamics is limited to transitions from the ground $X 0^+_g$ state to the first dissociation limit $I(^3P_{3/2})+I(^3P_{3/2})$ via states such as the $A1_u,C(B^{''})1_u, B^{'}(0_u)$, or to vibration motion via excitation to the bound $B 0^{+}_u$ state. Upon absorbing two photons, non-resonant and resonant Raman processes can take place to excite a vibration wavepacket at the ground state via $X\leftarrow M \leftarrow X$  type transition with $M$ the allowed symmetry intermediate state for the resonant case. Furthermore, two-photon absorption can lead to dissociation at all the dissociation limits, for example by exciting to the $C 1_g$ and $0^+_g$ states. At three-photon absorption, excitation of Rydberg and IP states become accessible, nonlinear Raman processes can take place to excite and mix various lower energy states, and the ionization threshold is reached via four-photon absorption

  The excited rovibrational motion will undergo rotational dephasing that will limit the delay time window such induced anisotropy 
  can be detected, while prompt dissociation along a single dissociative state will preserve the anisotropy. 
  
Here, we present the first results for the case of a strongly driven molecular Iodine vapor, where multiphoton processes take place, beyond the perturbative single or two-photon absorption processes.  We show that we can analyze and retrieve time-dependent high-fidelity high-order anisotropy information of an ultrafast X-ray scattering signal and assign motions of various excitation processes that take place simultaneously de-novo.    

\section{Theory}
In this section we shall discuss how to link between the spatial-temporal information of a photoexcited  molecular system from a wavepacket perspective, with its observed scattering pattern, using some results that have been derived  similarly before in other studies \cite{Ben-Nun_Cao_Wilson_1997,Lorenz_Moller_Henriksen_2010,lorenz2010interpretation}.  
We start with the assumption that we have solved a time-dependent Schrodinger equation (TDSE) and have the exact charge density of an excited system. We express the charge density by an optically excited nuclear wave packet that is propagated along some electronic state. 

We then obtain expressions for the scattering pattern this wavepacket generates. We will be interested in the general case of an n-photon absorption that induces a non-trivial anisotropy and will result in up to $2n$ order Legendre polynomials terms in the scattering signal.  
We also simplify the treatment assuming both the X-ray and optical laser pulse are co-propagating along $\vec k_0$, and that the angle between the incident X-ray beam and the laser polarization is $\pi/2$, allowing the Legendre decomposition approach in angle space. Scattering from time-evolving systems is generally inelastic \cite{Lorenz_Moller_Henriksen_2010,simmermacher2019electronic}, however, assuming typical experimental conditions, we can use the so-called static approximation \cite{Lorenz_Moller_Henriksen_2010} and replace the general electronic scattering operator with its elastic expression for the differential cross-section for scattering:  
 
\begin{equation}
\frac{d\sigma}{d\Omega} = \sigma_T   \int d\vec R \rho(\vec R,t) |F(\vec q,\vec R)|^2  \label{eq:crosssec1}
\end{equation}

where $\sigma_T$ is the Thomson cross-section, $\rho$ is the charge density,  and the squared molecular form factor $|F|^2$ is invariant under space inversion $R \mapsto -R$.  If we further simplify the treatment and consider only a single atom pair, as in the case of diatomic Iodine,  we can then write the differential cross-section:

\begin{equation}
\frac{d\sigma}{d\Omega} =  \sigma_T f_1(q) f_2(q) \int d\vec R \rho(\vec R,t) (e^{\imath \vec q \vec R} +e^{-\imath \vec q \vec R})  \label{eq:crosssec2}
\end{equation}

with $f_i(q)$ the i$^{th}$ atomic form factor. We can expand the scattering exponential term using the plane wave expansion:
\begin{equation}
e^{\imath \vec q \vec R} = \sum_{n=0,2,\ldots} (2n+1)(-1)^{n/2} P_n(cos \theta_{qR}) j_n(qR)
\end{equation}
 
where $P_n$ are Legendre polynomials, $j_n$ are spherical Bessel functions, and $\theta_{qR}$ is the angle between $\vec q$ 
and $\vec R$.  
We only need to sum over the even orders as the components with odd polynomials are anti-symmetric under space inversion and will cancel when we will calculate the differential cross-section. 

We use the spherical harmonics addition theorem to expand the Legendre polynomials to express $\theta_{qR}$ in terms of the experimentally measured scattering angles $(\theta_q,\phi_q$),  and $(\theta,\phi)$, the angle between the laser polarization and molecular axis and its corresponding azimuth:

\begin{equation}
P_n(cos \theta_{qR}) = \frac{4 \pi}{2n+1}\sum\limits_{j=-k}^k Y_{jk}^*(\theta_q,\phi_q) Y_{jk}(\theta,\phi)
\end{equation}

Using the scattering symmetry, we can integrate over $\phi_q$ and use the expressions above to arrive to:

\begin{equation}
\frac{d\sigma}{d\Omega} = 4 \pi \sigma_T  \sum_{n=0,2,\ldots}  (-1)^{n/2} P_{n}(\cos \theta_{q}) S_n(q) 
\label{eq:diffraction_pattern}
\end{equation}

\begin{equation}
\begin{split}
S_n(q) =   f_1(q) f_2(q) \int_{-1}^{1} d \cos(\theta)   \int_{0}^{\infty} dR   \int_{-\infty}^{\infty} dt  \Xi(t-\tau)  R^2  \\
|\psi(R,\cos(\theta),\tau)|^2 P_{n}(\cos(\theta)) j_n(qR) 
\end{split}
\label{eq:Sn}
\end{equation}

where $\Xi(\tau)$ is the X-ray pulse intensity envelope, and $\psi(R,\cos(\theta),\tau)$ is the molecular wavepacket. The n$^{th}$ order process will be manifested both by the intensity distribution on the detector via the scattering angle $\theta_q$, and via the anisotropy curves $S_n(q)$. A schematic description of the relevant coordinates and experimental approach is described in Fig \ref{fig:Scheme}.   The subsequent analysis of the scattering pattern given by Eq \ref{eq:diffraction_pattern}  will use a Legendre decomposition over the detector angle $\theta_d$ to  recover $S_n(q)$.

\section{Methods}

The experimental procedure is described in detail in a previous study\cite{glownia2016self}. In short, time-resolved X-ray scattering was performed at the Linac Coherent Light Source (LCLS) free-electron laser (FEL) facility, SLAC National Accelerator Laboratory, using the hard X-ray pump-probe (XPP) instrument. Molecular Iodine vapor with a column density of $\sim 10^{18} cm^{-2}$ was excited by a  520 nm, 40 $\mu$J, 50 fs, optical pulses of $\sim 5 \times 10^{11} W/cm^2$, and probed by a 9 keV, 2 mJ, 40 fs X-ray pulses at a variable pump-probe time delay provided by the LCLS. Approximately $10^7$ X-ray photons per pulse were scattered onto a 2.3 megapixel array detector \cite{hart2012cspad}, with $\sim 50$ photons per pulse per pixel. 

Initial processing of the raw scattering data included detector corrections that are mentioned in a previous study \cite{glownia2016self}.  The analysis includes a "dark" detector correction, where an average image of the detector without incident x-rays is subtracted from the raw data, following single-pixel detector corrections due to their non-linear response. The images are then corrected for polarization\cite{hura2000high} of the LCLS pulses, as well as scattering geometry for the case of a plane detector \cite{bosecke1997small}.  Subsequently, the scattering patterns were corrected for absorption artifacts due to the scattering cell geometry and upstream beam noise by imposing an isotropy condition\cite{van2016atomistic} that is expected for the unexcited scattering signal from Iodine thermal ground state. 
The images were sorted according to a jitter correction timing tool and averaged to obtain a pump-probe delay resolution of 20 fs.  While the resolution of such time delay binning exceeds the excitation pulse duration, it is appropriate for the case of the multi-photon excitation observed which has a shorter effective duration, as well as to the jitter correction resolution \cite{harmand2013achieving,glownia2019pump}. We improve the signal to noise ratio by analyzing the data from several pump-probe scans that are combined and sorted according to the pump-probe time delay.  

\begin{figure*}[hbt!]
	\includegraphics[width=1\textwidth]{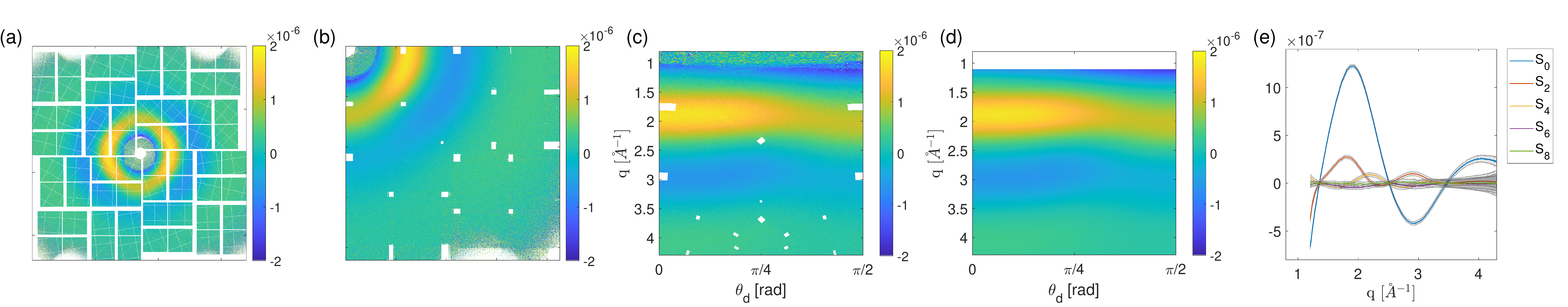} 
	\caption{(a) The LCLS 2.3 megapixel array detector (CSPAD \cite{hart2012cspad}) showing the difference scattering signal $\Delta I^{(\tau)}=I^{(\tau)}-\langle I^{\tau<0} \rangle $ for a pump-probe delay of $\tau= 240 fs$. (b) The scattering signal is four-folded and averaged and (c) transformed to polar detector coordinates. The usable  $q$ range for analysis was $1.2\angstrom^{-1}<q<4.3 \angstrom^{-1}$ due to the scattering cell geometry and mask used. (d) Applying Legendre decomposition to the polar representation captures its anisotropy orders as well as filters for higher angle dependent noise and missing signal. The reconstruction of the polar signal here was done using Eq. \ref{eq:Leg_decomp} where we used Legendre orders up to $P_{12}$. (e) We use the Legendre coefficients to obtain the anisotropy curves $S_n$ (Eq. \ref{eq:Beta2Sn}) and show the estimated standard error per curve (shaded areas).}
	\label{fig:rawdata}
	
\end{figure*} 

A representative time-binned difference signal is shown in Fig \ref{fig:rawdata}a.  We use the cylindrical symmetry of the scattering signal to increase its fidelity without loss of angle resolution by four-folding and averaging the scattering image quadrants. The averaging is weighted by the number of detector pixels that contribute from each quadrant. Four-folding the detector image also helps reduce the effective missing data that appears as gaps between the ASIC elements of the array detector. The remaining missing data points in  Fig \ref{fig:rawdata}b are excluded from the later Legendre decomposition analysis. The signal is then transformed and binned in the ($q$,$\theta_d$)  polar coordinates.

We decompose the signal in each $q$ bin to even order Legendre basis up to the relevant significant order (Fig \ref{fig:rawdata}d):
\begin{equation}
I(q,\theta_d) =   \beta_0 (q) \sum_{n=0,2,\ldots} \widetilde{\beta}_n(q)  P_{n}(\cos \theta_{d})   
\label{eq:Leg_decomp}
\end{equation}

With the radial intensity $\beta_0(q)$ and the normalized detector anisotropy terms $\widetilde{\beta}_n(q) = \beta_n(q) / \beta_0(q)$. The fit is done only on $\theta_d$ values that contain signal.  The relation between the n$^{th}$ order anisotropy curve $S_n(q)$ in Eq. \ref{eq:Sn} and the corresponding  $\beta_n(q)$ term is given by
\begin{equation}
S_n(q) =    \frac{\beta_0(q) \widetilde{\beta}_n(q)} {\cos^n(\theta_q)}=  
\frac{\beta_n(q) }  { \left(1-  \frac{q^2}{4|k_0|^2}   \right)^{n/2}}  \text{,    n=0,2,\ldots}
\label{eq:Beta2Sn}
\end{equation}
where $|k_0|$ is the length of the wave vector of the incoming X-ray beam.
We note that while $\beta_0(q,\tau)$ has units of intensity or the average number of scattered photons per $q$, the higher-order $\widetilde{\beta}_n(q,\tau)$ are dimensionless and represent ratios between the relevant angle components that dictate the degree of anisotropy.

Additional care is needed in centering the images before they are being four-fold averaged and transformed to polar coordinates. The residual error to $S_n$ due to centering inaccuracy was simulated for the detector binning used in the analysis. For example, in the case of a pure isotropic signal $S_0$, we find that a centering error at the single bin level $dq \sim 4 \times 10^{-3} \angstrom^{-1}$ will create an artificial residual signal for the higher-order $S_{n>0}$ that amounts to $\sim 10^{-4} S_0$  across the $q$ range used.  Deviation from the correct image center beyond the single bin resolution will create artificial anisotropy that will obscure and distort the weak signals of the high anisotropy orders.  
  
We find the center of the images by analyzing time delays where scattering can only be isotropic and validate our approach by limiting its residual anisotropy below the single-bin limit discussed. We analyze the contrast signal:
\[ I_{c} = \frac{\langle I^{(\tau<0)}\rangle-\langle I^{(\tau \gg 0)}\rangle}{\langle I^{(\tau<0)}\rangle+\langle I^{(\tau \gg 0)}\rangle} \]
Where $\langle I^{(\tau<0)}\rangle$ is obtained by averaging the scattering signal before the laser excitation, containing only isotropic scattering, and  $\langle I^{(\tau \gg 0)}\rangle$  is the averaged signal at time delays after rotational dephasing took place and most the dynamics have equilibrated.  We then apply to $I_c$ a series of intensity thresholds, each yields a scattering signal $\tilde{I_c}$ that will be distributed symmetrically around the center. We find the centers of these distributions using a Random Sample Consensus (RANSAC) algorithm \cite{torr2000mlesac},  an iterative method adapted to robustly fit circles in the presence of noise.  

Briefly, data points in $\tilde{I_c}$ are classified as outliers or inliers, and the fitting procedure ignores the outliers.  The classification is done by randomly sampling a small subset of points to estimate the model parameters. The condition for points to be  considered inliers is given by $|\sqrt{(x_i-x_c)^2+(y_i-y_c)^2}-r|<d$, where $x_i,y_i$ are data points of the subset , $x_c, y_c, r$ are the estimated model parameters (center and radius), and $d$ is the distance fit tolerance.  The random sampling repeat until the fraction of inliers over the total number of points that share the same model parameters exceeds a threshold.
We then take the trimmed mean of the estimated centers of $\tilde{I_c}$  as the center of the scattering image, and obtain for the case of isotropic signal, a residual $<10^{-4} S_0$ for the higher-order $S_{n>0}$, indicating centering accuracy at the single bin level.

The values and estimated standard errors for the measured anisotropy curves $S_n(q)$ are shown in Fig \ref{fig:rawdata}e. We calculate the standard error by first measuring the experimental weighted sample variance of each $\Delta I(q,\theta_d)$ element at each time bin delay. The variance for the $k^{th}$ time delay difference signal $\Delta I^{(k)}(q,\theta_d)$ is obtained by:   
\[
Var(I^{(k)})=\frac{1}{N^{(k)}} \sum_{m=1}^{N^{(k)}} |w(q,\theta_d)  \big(\Delta I^{(k)}(q,\theta_d)-\Delta I^{(k)}_m(q,\theta_d) \big)|^2
\]

where $N^{(k)}$ is the number of images recorded for that time delay bin, $w(q,\theta_d)$ is the statistical weight each element in $(q,\theta_d)$ has due to the four-folding and detector pixel binning, $\Delta I^{(k)}= I^{(k)}-\langle I^{\tau<0} \rangle$ is the weighted average intensity difference of the $k^{th}$ time bin, and $\Delta I_m$ expresses the $m^{th}$ image that belongs to the $k^{th}$ time bin.  The measured variance is then translated to a weights vector in the weighted least squares Legendre fitting process. The standard error obtained from the fit is calculated for each $q$ bin and is propagated to each anisotropy order according to Eq. \ref{eq:Beta2Sn}. 

Because we will be measuring a time-dependent signal $S_n(q,\tau)$ we would like to use its anisotropy information as a function of the temporal pump-probe delay to recover and disentangle the different multiphoton processes it captures.  An approach we have recently introduced to successfully characterize one and two-photon interaction of isotropic scattering signals \cite{bucksbaum2020characterizing,ware2019limits} employs a temporal Fourier-transform to obtain frequency-resolved X-ray-scattering signals. 
In this approach, time-periodic vibration motions in $S(q,\tau)$ will appear as peaks in the frequency domain $S(q,f)$ , while ballistic dissociation  motions will appear as lines where the dissociation velocity is linearly proportional to $v=2 \pi f/q$, where $f$ is the Fourier-transformed frequency coordinate.

The reason for such behavior can be understood if we model motion by a simple outgoing charge density described by $\delta(R(t))=\delta(R-v\tau)$. The time dependence in the scattering signal is then weighted by the spherical Bessel function according to the $qR(t)$ product. The Fourier transform of such function $\int d\tau e^{\imath \omega \tau} j_n(qR(\tau)) $ will result in an exponential integral function that has a maxima when $qR-\omega R/v=0$. A detailed derivation of this result can be found in Ref \cite{ware2019characterizing}.  We shall discuss implementing such an approach in recovering the anisotropy information for the case of multiple order contributions.

\begin{figure*}[hbt!]
	\includegraphics[width=1\textwidth]{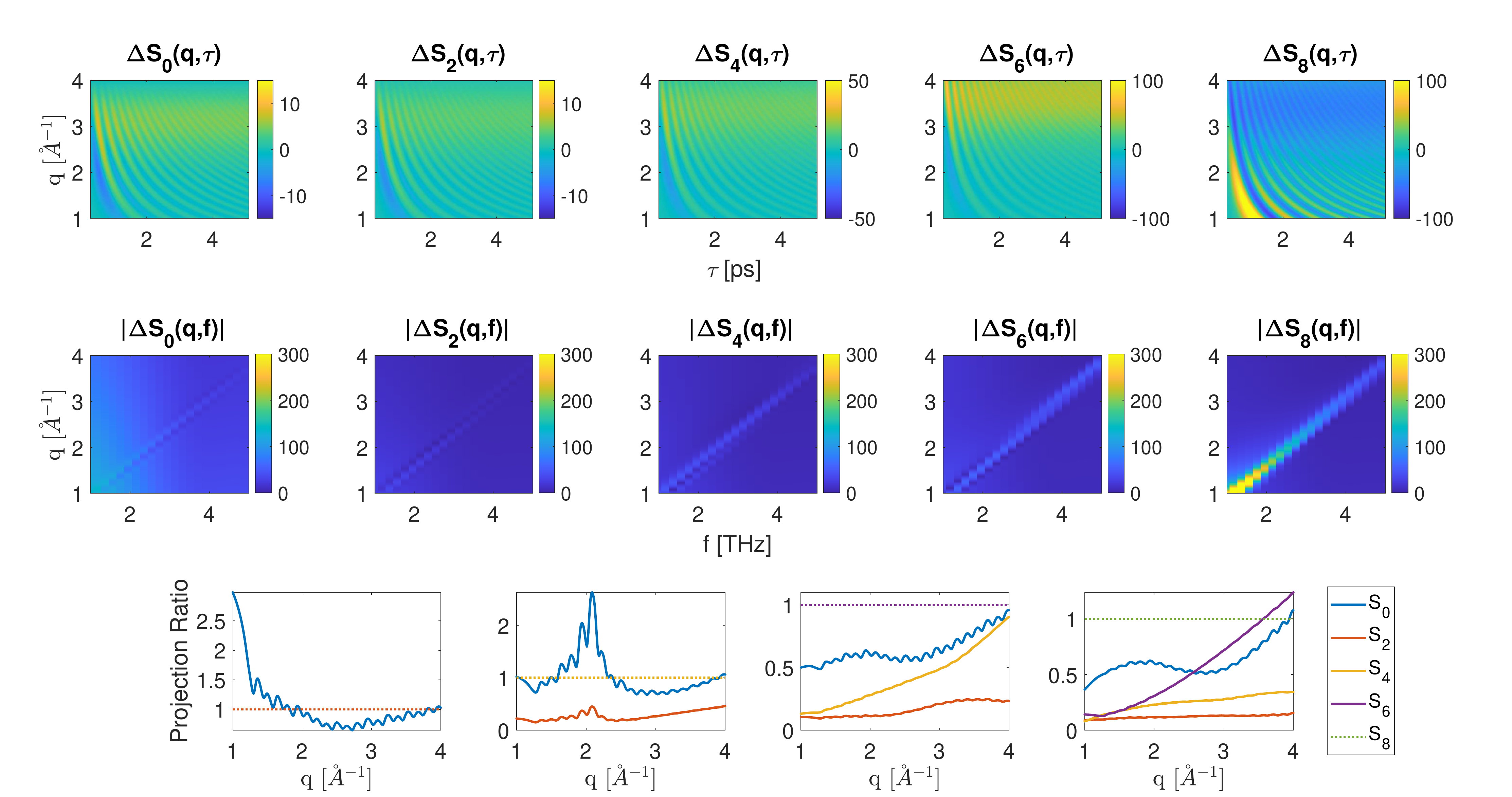}
	\caption{(top) Simulated difference signal of the anisotropy curves for the case of dissociation with a $cos^8(\theta)$ angular distribution. The highest order of anisotropy that is obtained is $S_8$, this term also captures the strongest signal in the $(q,\tau)$ domain.  However, the angular distribution modeled has also projections on all lower $S_{k<n}$ orders. (middle) Fourier transforming the temporal domain and plotting the magnitude of each order reveals the dissociation signature as straight lines along $2 \pi f = q v$.  (bottom) Analysis of how the dissociation signal of order $n$ propagates to lower orders $k<n$ as a function of $q$.  In each plot, we normalized the $k<n$ order signal by the leading order $n$  (marked dotted line at unity). The most right-side plot shows the case for $n=8$ that is obtained from the middle row of the figure. This type of signal propagation mapping from higher to lower orders allows tracing the origins of each anisotropy for the general case of multiple order contributions.}
	\label{fig:SN_theory_tau2f}
\end{figure*}

Consider an anisotropic dissociation motion modelled by the charge density:

\[ |\psi(R,\cos(\theta),\tau)|^2= \delta(R-(R_0+v\tau)) \cos^{n}(\theta) \]
where $R_0$ is the initial position at the instance of dissociation, $v$ is the velocity of the dissociation of anisotropy order $n$. We use Eq. \ref{eq:diffraction_pattern}  and Eq. \ref{eq:Sn} to simulate the scattering pattern on a detector at each pump-probe delay and truncate the scattering signal in the $q$-range similar to typical experimental conditions, $1 \angstrom^{-1} < q <4 \angstrom^{-1}$.  We subtract the time delayed scattering signal from the stationary signal $\Delta I =I(q,\tau)-I(q,\tau=0)$,  and decompose the difference signal in Legendre polynomials  to obtain the $\widetilde{\beta}_n(q,\tau)$ coefficients as shown in Eq \ref{eq:Leg_decomp}. We then obtain the difference anisotropy curves $\Delta S_n(q,\tau)$ using Eq. \ref{eq:Beta2Sn}, and Fourier transform them to obtain the dissociation signature in $(q,f)$ space.  

For example, in Fig \ref{fig:SN_theory_tau2f} we demonstrate the case for a pure anisotropy of order $n=8$.  We present the time-dependent anisotropy difference curves  $\Delta S_n(q,\tau)$ and their temporal Fourier transforms. We verify that we obtain up to the 8$^{th}$ order Legendre coefficients without any higher orders contributing. We also observe that the highest anisotropy order measured  ($\Delta S_8$ in the example) has the strongest contribution both in the both temporal and frequency domains. The analysis captures the expected behavior where the dissociation velocity is obtained via $v=2\pi f/q$,  however, we also note that the same dissociation signature is present at lower anisotropy orders. 

The reason for the appearance of a similar dissociation line in the lower orders is because of the nature of the Legendre polynomials used in the decomposition. A scattering signal from a $cos^n$ distribution will not only be captured by $P_n$ order polynomial but have projections to all $k<n$ order Legendre polynomials. In this example,  a $cos^8$ distribution will be decomposed to a Legendre series of even orders up to $n=8$. For a general case where several orders contribute, one may choose to only analyze the $n=2$ order signal instead of all orders. This order includes besides the $cos^2$ contribution the projections of all higher orders, thus obtaining a representation of the total anisotropy in the sample. Doing so, however,  hinders the assignment of specific processes with a particular degree of anisotropy, as well as attenuates higher-order contributions by at least an order of magnitude.

Instead, we show we can map the way the signal of each anisotropy order is projected among lower orders to trace how different order contributions are observed given typical experimental sampling in $q$ and $\tau$. In Fig \ref{fig:SN_theory_tau2f} we also show a calculation of the expected projection as function of $q$ for orders up to $n=8$. We normalize the projections by the leading order and obtain typical ranges of the magnitude of signal propagation, as summarized in Table \ref{tab:table1}. We will use the information of that mapping to disentangle dissociation processes among the orders and uncover the contribution of each order.

\begin{table}[h]
\small
	\caption{\label{tab:table1} The ratio ranges of $\mathcal{P}|S_k(q,f)| / |S_n(q,f)|$, where $\mathcal{P}|S_k|$ is the $k^{th}$ order projection  of $|S_n(q,f)|$ with $k=0,2,..n-2$. The numbers in each cell present the value limits obtained for the range $1 \angstrom^{-1}<q<4 \angstrom^{-1}$.  The values in the table are obtained from the curves seen at the bottom row of Fig \ref{fig:SN_theory_tau2f} .  }
 
		  \begin{tabular*}{0.48\textwidth}{@{\extracolsep{\fill}}c|c|c|c|c}
			\hline
			& $\mathcal{P} S_0$ & $\mathcal{P} S_2$ & $\mathcal{P}S_4$ & $\mathcal{P}S_6$   \\
			\hline
			$S_2$  &  0.65 - 2.98  &    1  & - & -   \\
			\hline
			$S_4$  & 0.68 - 2.62   &    0.15 - 0.46  &   1 & -  \\
			\hline
			$S_6$ & 0.49 - 0.96  &   0.09 - 0.25  &  0.13 - 0.9  &  1     \\
			\hline
			$S_8$ & 0.36 - 1.07  &  0.09 - 0.15   &  0.08 - 0.34   &  0.13 - 1.23     \\
			\hline
		\end{tabular*}
 
\end{table}

\section{Results and discussion}

\begin{figure*}[hbt!]
	\includegraphics[width=1\textwidth]{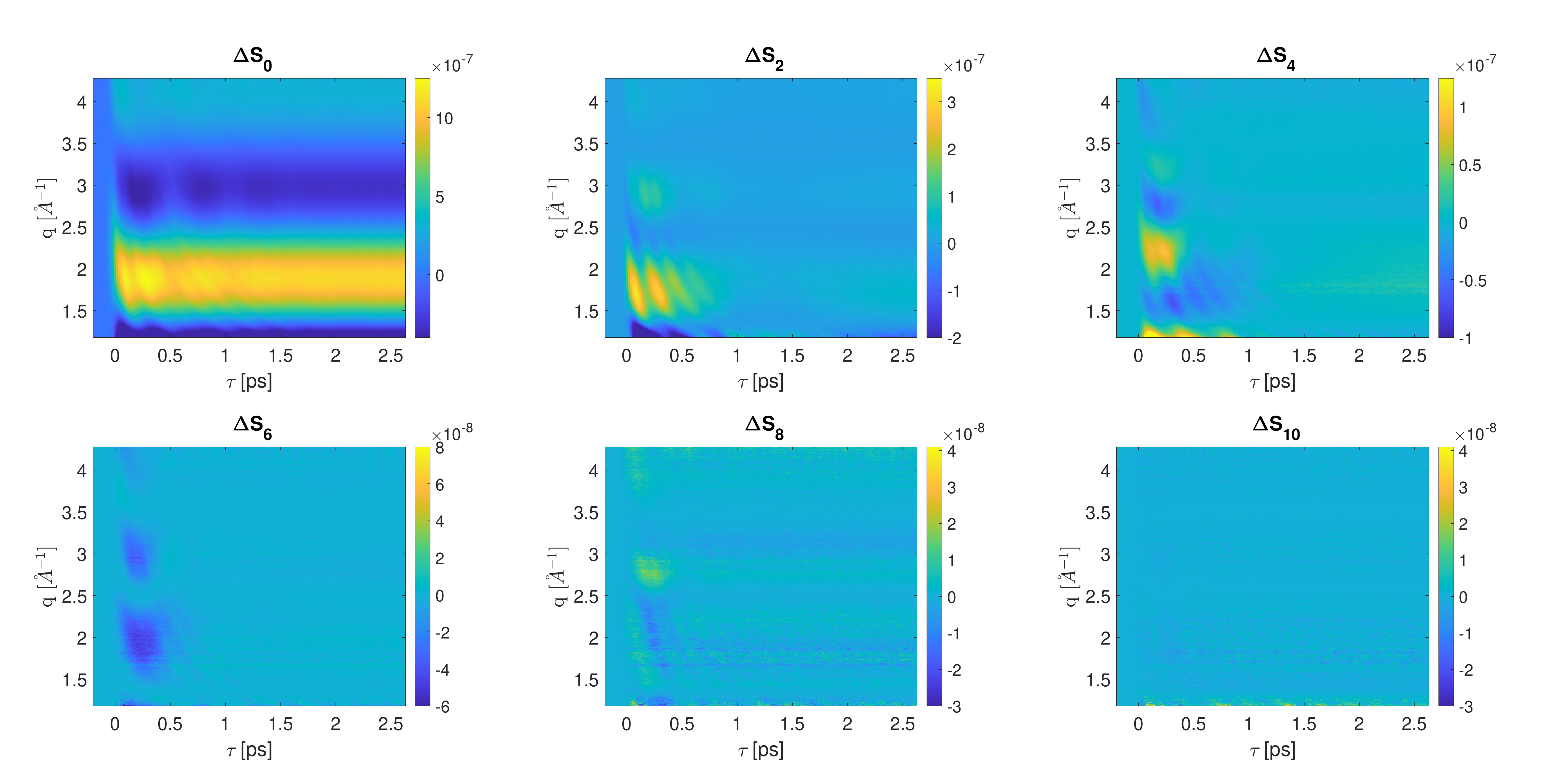}
	\caption{The experimental anisotropy curves $\Delta S_n$ obtained from Legendre decomposition of the measured scattering signal.  We observe anisotropy up to $n=8$ order. The $n=0$ term is the angle averaged scattering signal that captures all dynamics. Time-dependent oscillations of several frequencies across the entire $q$ range can be seen (most noticeable at $2<q<3$), caused by various single and multiphoton vibration excitations. Higher $\Delta S_n$ terms capture various dissociation processes that are manifested by $q$-dependent modulations curves, the longer the dissociation distance the faster the modulation in $q$.  The $q$-dependence nature of the amplitude in the various orders is due to the spherical Bessel $j_n(qR)$  dependence as seen in Eq. \ref{eq:Sn}. The drop of signal in the higher-order terms as a function of delay is due to rotational dephasing for the case of the molecules that are excited to bound states. However, excitations that lead to dissociation carry the anisotropic nature of the signal regardless of dephasing as the atoms produced in the dissociation process will move outward along the straight line defined by the internuclear axis of the molecule.}
	\label{fig:betas}
\end{figure*} 

We have applied the Legendre decomposition as described in Eq. \ref{eq:Leg_decomp} to the measured scattering difference signal $\Delta I^{(\tau)} = I^{(\tau)}(q,\theta_d)-\langle I^{(\tau<0)}(q,\theta_d) \rangle$ to derive the experimental anisotropy curves $\Delta S_n(q,\tau)$ using Eq. \ref{eq:Beta2Sn}. The result is seen in  Fig \ref{fig:betas} for the different anisotropy orders. The decomposition was done up to order $n=12$ and we have found that for the experimental conditions used it was sufficient to only consider orders of up to $n=8$, indicating that multiphoton processes up to 4-photon absorption are taking place.  The signal at $\Delta S_{10}(q,\tau)$ has a negligible contribution for the analysis, as its observable signal at $q\sim 1.2 \angstrom^{-1}$ does not allow information to be extracted, and may be related to the limit of anisotropy detection at this range due to the asymmetry of the inner hole mask of the detector. 

The entire excitation dynamics are captured in the angle averaged $\Delta S_0(q,\tau)$ term, where we primarily observe two types of signals: Time periodic oscillations across the entire $q$ range related to various vibrational excitations, and, $q$-dependent modulations that change their radius of curvature as a function of time and indicate dissociative motion.  The dissociation signature is also evident in the higher 
$\Delta S_n(q,\tau)$ terms, as this type of signal is not sensitive to rotational dephasing where fragmentation preserves the molecules axis.

\begin{figure*}[hbt!]
	\includegraphics[width=1\textwidth]{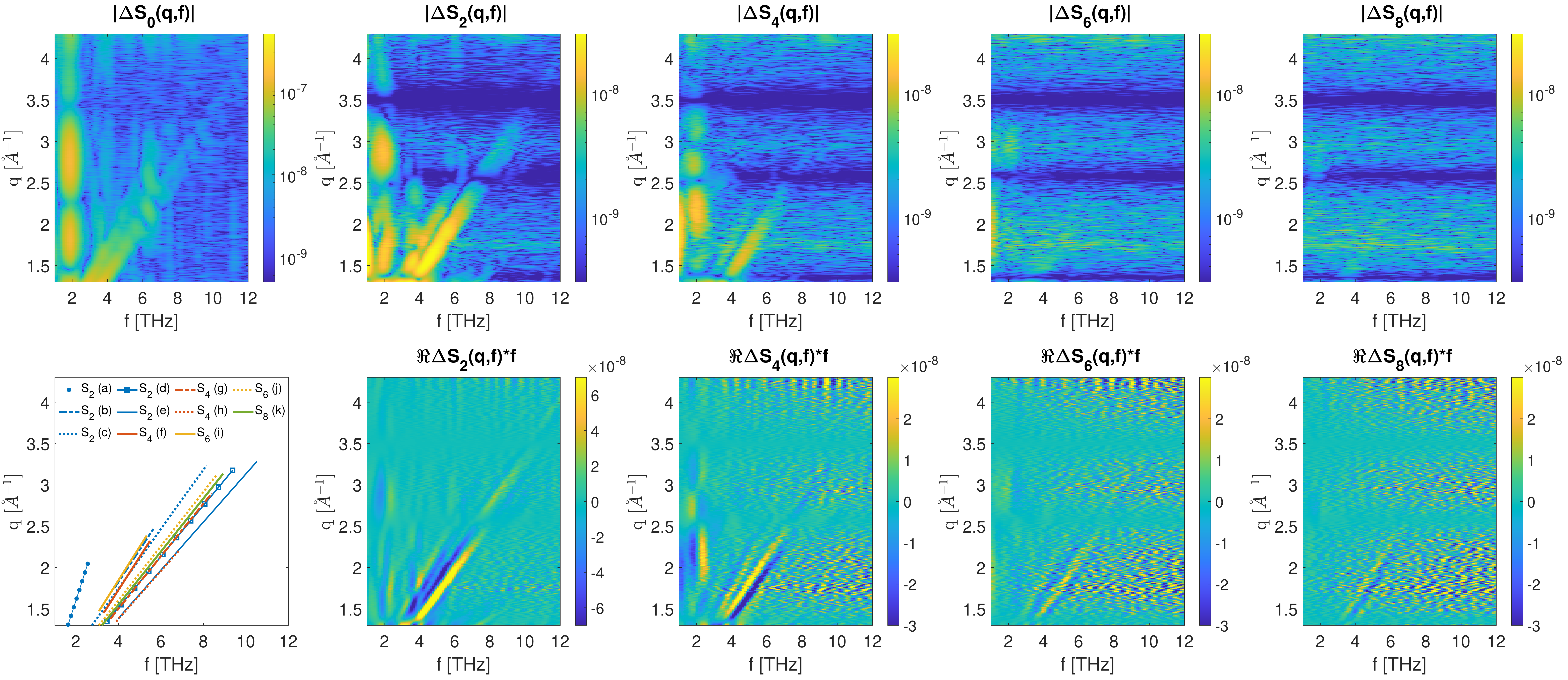}
	\caption{The (top) magnitude and (bottom) real part of the measured Fourier transformed anisotropy curves $ \Delta S_n(q,f) $. For visualization we use a logarithmic scale for the magnitude.  The $|\Delta S_0(q,f)|$ term captures all dynamics, including many multiphoton vibrationaly excited states that appear as peaks along $f$. These peaks broaden and attenuate in higher orders due to rotational dephasing that limits their time periodic sampling. Several dissociation pathways that appear as diagonal lines are captured in the $S_2$ and $S_4$ orders, as well as weaker dissociation in the $S_6$ and $S_8$ orders. To analyze the dissociation signatures we also use the information of the real or imaginary part of $S_n(q,f)$ that allows better contrast vs the noise floor, as seen for the terms $\Re{S_n(q,f)}$ scaled by the Fourier vector that appear below the magnitude information. (bottom left) Analyzing the positions of the dissociation signatures in different anisotropy orders allows  to identify the pathways and the number of photons that participated (see text). The line colors code for the different anisotropy orders, the slopes correspond to the average dissociation velocities:  (a) 7.9 (b) 16.1 (c)  17.5 (d)  19.1 (e) 21.6 (f)  15.7 (g) 19.8  (h)   21.6  (i)   15.5 (j) 20.4  (k) 19.6 $\angstrom /ps \pm 0.5 \angstrom /ps$.   }
	\label{fig:SN_f}
\end{figure*} 

We then Fourier transform the anisotropy curves to obtain the frequency-resolved $\Delta S_n(q,f)$ terms as seen in Fig \ref{fig:SN_f}. We analyzed the magnitude and real part of each term and obtained the estimated dissociation velocities for the different anisotropy orders via a linear model estimation using a RANSAC algorithm. For vibration frequencies detected, the estimate was based on the center of mass of the peak found in frequency.  We observe that the dissociation velocities take place at a range of $8-22 \angstrom/ps$. We analyze the signal strengths of each dissociation line starting from the highest order measured and consider the way the signal can propagate to lower orders to resolve its origin and the pathway that takes place. We also observe a multitude of vibration excitation that is manifested as peaks in frequency at a wide $q$ range, most noticeably in the $|\Delta S_0(q,f)|$ term.  Analysis of the vibrational excitation is limited in the high order anisotropy terms due to rotational dephasing that restricts the effective time-window to sample periodic motions. As a result, for Iodine, slower vibrations will only have two to three periods before the signal will be lost to dephasing. This broadens the detected peak widths and restricts resolving details in frequency.  

To consider the way the signal of higher-order anisotropy terms is projected to lower orders, we start by analyzing the highest observable order.  The observed dissociation signal of $|\Delta S_8(q,f)|$ indicates that a 4-photon transition to a repulsive \emph{gerade} excited state occurred. From the estimated dissociation velocity of $19.6 \pm 0.5 \angstrom/ps$ we calculate a kinetic energy release (KER) of $2.5 \pm 0.1 eV$ that corresponds to a transition to the $C 1_g$ state that dissociates to the second dissociation limit:  $I(^3P_{3/2})+I(^3P_{1/2})$. This transition is accomplished via a 3-photon absorption to Rydberg states following a 1-photon transition back to the  $C 1_g$ state, as well as a Raman excitation to the ground state following 2-photon absorption. Furthermore, 4-photon absorption from the ground state can populate the 6d-Rydberg series which lies above the dissociation limit of the first-tier IP states. While these states are known to give rise to several fragmentation channels, these  processes take place at a much later time compared to the time delays probed here. Previous studies that measured the fragmentation of these Rydberg states  \cite{donovan2015heavy,bogomolov2014predissociation} show that there was no anisotropy measured for the fragments. This result indicates that dissociation happened at a timescale longer than the rotation period of Iodine, at least an order of magnitude larger than the range of time delays probed here.  Around  the range  $1.5\angstrom^{-1} < q <2.5 \angstrom^{-1}$ where the dissociation signal was observed, we deduce that  $ \sim 30 \%$ of the signal in $|\Delta S_8(q,f)|$ is projected to the $|\Delta S_6(q,f)|$. Similarly,  the projections to $|\Delta S_4(q,f)|$ and $|\Delta S_2(q,f)|$ are about $20 \%$ and $10 \%$ respectively.

The observed dissociation signal of the  $|\Delta S_6(q,f)|$ term is related to a 3-photon transition to an \emph{ungerade} repulsive excited state, as well as to a possible projection from $n=8$ order. We measure two dissociation signatures for this term, with estimated  velocities of  $15.5\pm 0.5 \angstrom/ps$ and $19.1\pm 0.5 \angstrom/ps$. The slower dissociation velocity agrees with the expected KER of $\sim 1.6 eV$ for nonlinear-Raman transition to the $C(B^{''})1_u$ state that decays to the first dissociation limit $I(^3P_{3/2})+I(^3P_{3/2})$.
The second dissociation signal with $19.1 \angstrom/ps$ is interpreted as a projection of the $n=8$ order of similar velocity. To arrive at that conclusion we sampled a region of interest around $q \sim 2 \angstrom^{-1}$ where the dissociation signal is present and calculated the ratio of the median intensity $\bar{|S_6|}/\bar{|S_8|}$ for this range. We obtain a ratio of $0.42$, which is similar to the expected range of values for a projection from $n=8$ in that $q$ range. Considering the noise floor observed for these terms in Fig \ref{fig:SN_f}, we deduce that most, if not all the signal observed at that dissociation channel is actually due to a projection of the $n=8$ order.  We also observe peaks around $1.4\pm 0.3$ and $2.1\pm 0.3$ THz corresponding to periods of 710 and 470 fs respectively. The first peak can be interpreted  as a 3 photon absorption process to high lying vibration levels $v=278-290$ of the $D0^+_u$ IP state, as well as for the $v=136$ range of the  $F0^+_u$ IP state  \cite{hoy1990reinvestigation,hiraya1988vacuum}. However, the limited resolution of the approach used cannot resolve finer details. The second detected peak is interpreted as a vibration wavepacket that was excited via nonlinear Raman scattering to the bound  $B(0^+_u)$ state with an estimated period of 450 fs.

For the  $|\Delta S_4(q,f)|$ term we observe three dissociation channels with velocities of  $15.7\pm 0.5 \angstrom/ps$,  $19.8 \pm 0.5 \angstrom/ps$, and $21.6 \pm 0.5 \angstrom/ps$. This order captures 2-photon transitions to a \emph{gerade}  excited state along with projections of the higher orders. Analyzing the relative signal strengths we observe that high order projections are minor here. The measured dissociation of  $15.7 \angstrom/ps$ correspond to a 2-photon transition to the $O^+_g$ state that dissociate via the 3rd limit $I(^3P_{1/2})+I(^3P_{1/2})$ with a KER of $1.6 eV$. The next dissociation is more prominent and relates to excitation of the $C 1_g$ state via the 2nd limit $I(^3P_{3/2})+I(^3P_{1/2})$ with a KER of $2.6 eV$. Further, we measure another dissociation signal with a faster velocity of $21.6 \angstrom/ps$ that corresponds to a KER of $3.1 eV$ indicating that we excite the $a 1_g$ state that decay to the 1st limit  $I(^3P_{3/2})+I(^3P_{3/2})$. We observe  peaks in frequency at $1.75 \pm 0.2 $ THz, $2.3 \pm 0.2 $ THz corresponding to periods of 570 and 430 fs respectively. We also observe a weaker peak around $6 \pm 0.2$ THz which indicates Raman excitation of the ground state at $\sim 167$ fs period.  The lower peaks around $2$ THz can only be explained by higher-order projections that are more visible in lower orders, as will be elaborated for the $n=2$ order. 

The lowest anisotropy order $n=2$ contains multiple dissociation channels that coincide with the higher orders, as also shown in Fig \ref{fig:SN_f} summary plot. In the perturbative limit this order will capture 1-photon transitions to \emph{ungerade} excited states such as the bound $B(0^+_u)$ state, the repulsive $C(B'')1_u$ and $A1_u$ states.  Analyzing the signal strength of the different channels concerning possible projections of the higher orders we learn that the $n=4$ order contributes most of the signal measured along the same dissociation channels with two exceptions. We measure a dissociation line at the $16.1 \angstrom/ps$ that correspond to 1-photon transition to the $C(B^{''})1_u$ state.  Additionally, the slowest dissociation velocity we measure $7.9 \angstrom/ps$ is related to a transition to the lowest repulsive $A1_u$ state with a KER of 0.41 eV. Such slow dissociation is at the limit of detection using the frequency-resolved approach as the line that corresponds to it has a greater overlap with the different vibration peaks, as seen around 1.5-2 THz. 

The $n=2$ order also shows multiple strong peaks in frequency up to 6 THz, that relate to vibration excitation, besides the strong and broad peak at 1.9 THz that correspond to the vibration motion of the $B(0^+_u)$. 
We see this effect also for the $n=4$ order lower frequency peaks. The ability to resolve vibration modes in frequency depends on the way rotation motion is projected to the different anisotropy orders. As a result, projections of higher to lower orders for periodic motions depend on the rotation dynamics in addition to the mathematical projection.  This contrasts with  to the way dissociation signal projection propagates, as it is insensitive to rotational dephasing. Consequently, the $S_2$ term effectively samples more time delay bins compared to higher orders when considering periodic motions. The attenuation and broadening of periodic signals due to rotational dephasing in the higher orders restrict the projection analysis between the orders to the point it is impractical to assign excitation pathways based on anisotropy.

The isotropic signal  $|\Delta S_0(q,f)|$ combines all types of signals, and in particular, can be used to identify all peaks in frequency that corresponds to several Raman and nonlinear Raman excitations.  While it can be challenging to resolve some of these peaks because of the overlap they have with all the diagonal dissociation lines, we conclude that the spectroscopy of vibrational motion is mostly limited to the isotropic signal in the context of the frequency-resolved X-ray scattering analysis presented here. 

\section{Conclusions}

In conclusion, we are presenting the first results of the observation of high-order anisotropy, up to 8$^{th}$ order, in ultrafast x-ray scattering.  We present an analysis method that uses a Legendre decomposition and Fourier analysis on measured and simulated scattering signal. We map the way different anisotropy orders project signal to lower orders and use it to resolve and interpret the experimentally observed signals. We resolve many dissociation channels and vibration modes that are excited by multiphoton transitions and assign the processes based on the order of the anisotropy they appeared at. Leveraging high-order anisotropy information of the scattering signal is limited by the rotational timescale. In particular, it is limiting the ability to use it for Fourier-resolved x-ray scattering of vibration motions, implying that direct real-space approaches should be considered in this context.  However, during that rotational timescale, the anisotropy information provides a sensitive filter to differentiate and trace different excitation pathways that take place simultaneously. The information provided using this approach can be also applied in other analysis schemes, as well as open the possibility to serve as a prior when analyzing the isotropic time-dependent signal past the rotation timescale.

\section*{Conflicts of interest}
There are no conflicts to declare.

\section*{Acknowledgments}

This work was supported by the U.S. Department of Energy, Office of Science, Basic Energy Sciences, Chemical Sciences, Geosciences, and Biosciences Division. The experiment described was carried out at the Linac Coherent Light Source (LCLS) at the SLAC National Accelerator Laboratory. LCLS is an Office of Science User Facility operated for the U.S. Department of Energy Office of Science by Stanford University.

\bibliography{main}% Produces the bibliography via BibTeX using main.bib.

\end{document}